\documentclass[notitlepage]{article}

\usepackage[affil-it]{authblk}

\usepackage[T1]{fontenc}
\usepackage[utf8]{inputenc}
\usepackage{textcomp}
\usepackage{amsmath}
\usepackage{graphicx}
\usepackage{esint}
\usepackage[english]{babel}

\renewcommand\[{\begin{equation}}
\renewcommand\]{\end{equation}}

\begin{document}

\selectlanguage{english}%

\title{The effect of rough surfaces on Nuclear Magnetic Resonance relaxation
experiments}

\author{Matias Nordin\thanks{Electronic address: \texttt{mnordin@stanford.edu}; Corresponding author} ~and Rosemary Knight}
\affil{Department of Geophysics, Stanford University,397 Panama Mall, Mitchell
Building 101, Stanford, CA 94305-2210, USA}

\date{\today}

\maketitle

\begin{abstract}
Most theoretical treatments of Nuclear Magnetic Resonance (NMR) assume
ideal smooth geometries (i.e. slabs, spheres or cylinders) with well-defined
surface-to-volume ratios (S/V). This same assumption is commonly adopted
for naturally occurring materials, where the pore geometry can differ
substantially from these ideal shapes. In this paper the effect of
surface roughness on the $T_2$ relaxation spectrum is studied. By homogenization
of the problem using an electrostatic approach it is found that the
effective surface relaxivity can increase dramatically in the presence
of rough surfaces. This leads to a situation where the system responds
as a smooth pore, but with significantly increased surface relaxivity.
As a result: the standard approach of assuming an idealized geometry
with known surface-to-volume and inverting the $T_2$ relaxation spectrum
to a pore size distribution is no longer valid. The effective relaxivity
is found to be fairly insensitive to the shape of roughness but strongly
dependent on the width and depth of the surface topology.
\end{abstract}

\clearpage

\section{Introduction}
It is well established that Nuclear Magnetic Resonance (NMR) measurements
of diffusing spins can be used to probe the geometry of porous media.
The NMR response is sensitive to the surface to volume ratio of the
confining pore space~\cite{Allen1997,brownstein_importance_1979}.
For smooth, ideal shapes, this relationship provides a means of estimating
the pore size distribution~\cite{brownstein_importance_1979,Loren1970}.
Naturally occurring materials may however have a more complicated
pore structure than these ideal shapes, leading us to pose the question:
Can there be a direct link to pore size when the pore geometry is
not ideal? 

The standard approach in a NMR relaxometry experiment is
to measure the transverse (spin-spin) relaxation time of water protons
in a porous medium and estimate the relaxation time $T_2$~\cite{torrey_bloch_1956}.
In addition to the relaxation processes in the bulk fluid the protons
interact with paramagnetic sites at the pore surface, which increases
the relaxation rate~\cite{Kleinberg1994206,korb}. In this way, the relaxation
experiment is sensitive to the surface-to-volume (S/V) ratio of the
confining porous medium. Brownstein and Tarr realized that the complex
nature of the spin-surface interaction is well described by a Robin
boundary condition~\cite{brownstein_importance_1979}. The magnetization
$m$ in a pore is then described by the following Bloch-Torrey equation 

\begin{equation}
\frac{\partial m(r,t)}{\partial t}=D\nabla^{2}m\label{eq:bloch-torrey}
\end{equation}
subject to a Robin condition at the pore boundary $(\hat{n}\cdot\nabla+\rho)m=0$
where the scalar $\rho\geq0$ denotes the surface relaxivity and $D$
the diffusion coefficient of the spins. In an NMR relaxation experiment
the relaxation rate(s) are sought. Omitting the bulk fluid relaxation
properties we restate the classical result 
\begin{equation}
\frac{1}{T_{2}}\sim\rho\frac{S}{V}\label{eq:fast-diffusion-limit}
\end{equation}
which has been derived on multiple occasions and is true in the limit
of the time $t$ approaching zero and in the so-called fast-diffusion
limit depicted by $\rho R/D<<1$ where R denotes the pore size~\cite{MitraRelaxation,brownstein_importance_1979}.
Equation \ref{eq:fast-diffusion-limit} allows an estimate of the
pore size distribution from NMR measurements in cases where the pores
can be approximated by ideal shapes (i.e. slabs, spheres or cylinders)
where a simple relationship between $S/V$ and the pore radius can
be established. Experiments ~\cite{keating_laboratory_2014} and
numerical studies~\cite{muller-petke_nuclear_2015} suggest that
not only will a deviation from ideal shapes disrupt this relationship
but that surface roughness may also have an impact; no rigorous investigation
of this latter effect has been conducted. 

In this paper we show that
when the surface is rough (as is true for most naturally occurring
materials) the above expression is no longer directly translatable
to the pore radius and that the derived pore size can differ substantially
from the actual size. We demonstrate that in the presence of surface roughness
the spins still behave as being in a smooth pore, but subject to a different,
effective, surface relaxivity. We provide a means of calculating this
quantity, by introducing a magnetization rate coefficient describing
the magnetic dissipation over the rough surface. 

\begin{figure}[b]
\centering{}\includegraphics[width=0.6\columnwidth]{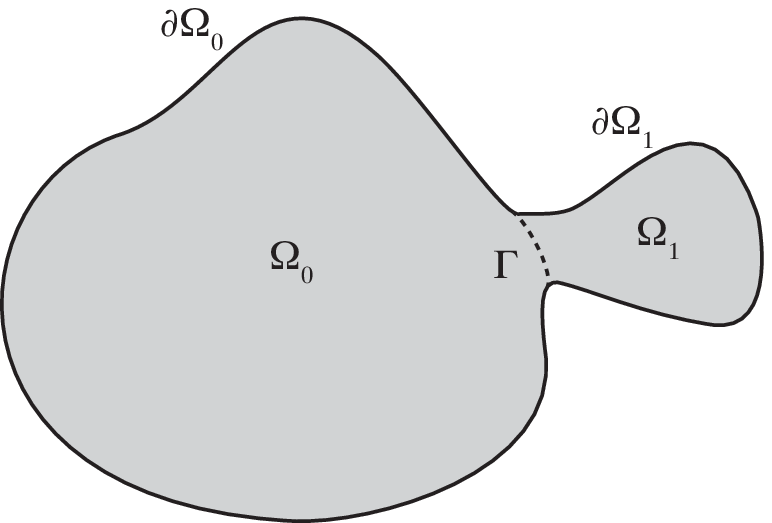}\protect\caption{A pore $\Omega_{0}$ with some a geometrical perturbation $\Omega_{1}$
separated by a fictitious boundary $\Gamma$.\label{fig:1} }
\end{figure}

\section{Theory}
We begin our analysis by considering a smooth pore $\Omega_{0}$ with
surface $\partial\Omega_{0}$ and with a small geometrical perturbation
$\Omega_{1}$, with surface $\partial\Omega_{1}$ (see figure \ref{fig:1})
representing surface roughenss. ``Small'' here means that the volume
(area in this 2D representation) of the perturbation is much smaller
than the volume (area in this 2D representation) of the original pore
$\Omega_{1}<<\Omega_{0}$. “Rough” means that the perturbation is
local with respect to the pore surface i.e. if we separate the two
domains with a fictitious boundary $\Gamma$ then the surface (length)
of $\Gamma$ is much smaller than the smooth pore surface (length)
$\partial\Omega_{0}$. Hence the geometrical perturbation $\Omega_{1}$
can be seen as a local roughening of the surface. We will utilize
this fictitious boundary below. 

Separating the diffusion equation depicted in equation \ref{eq:bloch-torrey}
yields the problem at hand: To find relaxation modes $u_{n}$ and
relaxation times $T_{n}$ satisfying
\begin{equation}
\begin{cases}
D\Delta u_{n}(r)=\frac{1}{T_{n}}u_{n}(r) & r\in\Omega\\
(D\hat{n}\cdot\nabla+\rho)u_{n}=0 & r\in\partial\Omega
\end{cases}\label{eq:bt}
\end{equation}
where $\hat{n}$ denotes the outward pointing normal of the total
pore surface $\partial\Omega$ and where $\Omega$ denotes the total
pore $\Omega=\Omega_{0}\bigcup\Omega_{1}$. We now proceed by considering
the right-hand side of equation \ref{eq:bt} as a charge distribution
$\tilde{u}$. The equations for the two sub-domains then become two
coupled electrostatic problems 
\begin{equation}
\begin{cases}
\Delta u_{i}(r)=\tilde{u} & r\in\Omega_{i}\\
(\hat{n}\cdot\nabla+\rho)u_{i}=0 & r\in\partial\Omega_{i}\\
u_{i}=u_{\Gamma} & r\in\Gamma
\end{cases}\label{eq:poissons_equation}
\end{equation}
for $i=0,1$ and some unknown charge distribution $u_{\Gamma}$ at
the boundary $\Gamma$. These equations are satisfied when the charge
distribution $\tilde{u}$ equals the original relaxation modes $\frac{1}{T_{n}}u_{n}(r)$
in equation \ref{eq:bt}. A general solution to the coupled electrostatic
problem in equation \ref{eq:poissons_equation} is given by 
\[
u_{i}=H(r)+G(r)
\]
 where $H(r)$ satisfies the Laplace equation $\nabla^{2}H(r)=0$
with the inhomogeneous boundary condition at $\Gamma$: $H(r\in\Gamma)=u_{\Gamma}$
and where $G(r)$ satisfies the Poisson's equation $\nabla^{2}G(r)=\tilde{u}$
with all homogeneous boundary conditions i.e. $G(r\in\Gamma)=0$.
By defining $\hat{n}\cdot\nabla G(r\in\Gamma)=\beta(r\in\Gamma)$
one may note that the solution $u_{1}$ satisfies the following inhomogeneous
Robin condition 
\begin{equation}
\hat{n}\cdot\nabla u_{1}+\alpha u_{1}=\beta\label{eq:inhomogeneous_rbc}
\end{equation}
at $\Gamma$ where we call the parameter $\alpha$ the magnetization
transfer coefficient 
\[
\alpha=\frac{\hat{n}\cdot\nabla H(r)}{H(r)}\mid_{r\in\Gamma}.
\]
This parameter describes the magnetic dissipation over the fictious
boundary $\Gamma$. The parameter $\beta$ may be estimated by noting
that the relaxation times are proportional to the pore volume, i.e.
\[
\frac{1}{T_{n}}\sim\frac{1}{\Omega_{0}+\Omega_{1}}\approx\frac{1}{\Omega_{0}}.
\]
Hence, by forming a Gaussian surface one may determine that
\[
\intop_{\Gamma}\beta(r)dr=\intop_{\Gamma}\hat{n}\cdot\nabla Gd\Gamma\sim\frac{\Omega_{1}}{\Omega_{0}}
\]
with $\beta$ approaching zero as $\Omega_{0}>>\Omega_{1}$. Therefore
the inhomogeneous boundary condition described by eq. \ref{eq:inhomogeneous_rbc}
approaches a homogeneous B.C. when the perturbation $\Omega_{1}$
is much smaller than the original pore. By continuity the same boundary
condition must hold in the larger pore $\Omega_{0}$ as well. The
magnetic transfer coefficient $\alpha$ can be estimated
by assuming that it is constant over $\Gamma$ $\alpha=c$ (a good
approximation when $\frac{\Gamma\rho}{D}<<1$ due to the form of the
relaxation modes). We then have
\[
\begin{aligned}\alpha_{a}=\frac{1}{\Gamma}\intop_{\Gamma}\frac{\hat{n}\cdot\nabla H(r)}{H(r)}dr=-\frac{1}{\Gamma}\intop_{\Gamma}\frac{\hat{n}\cdot\nabla H(r)}{c}dr=\\
-\frac{1}{c\Gamma}\intop_{\partial\Omega_{1}}\rho H(r)dr.
\end{aligned}
\]
\begin{figure}
\centering{}\includegraphics[width=0.8\columnwidth]{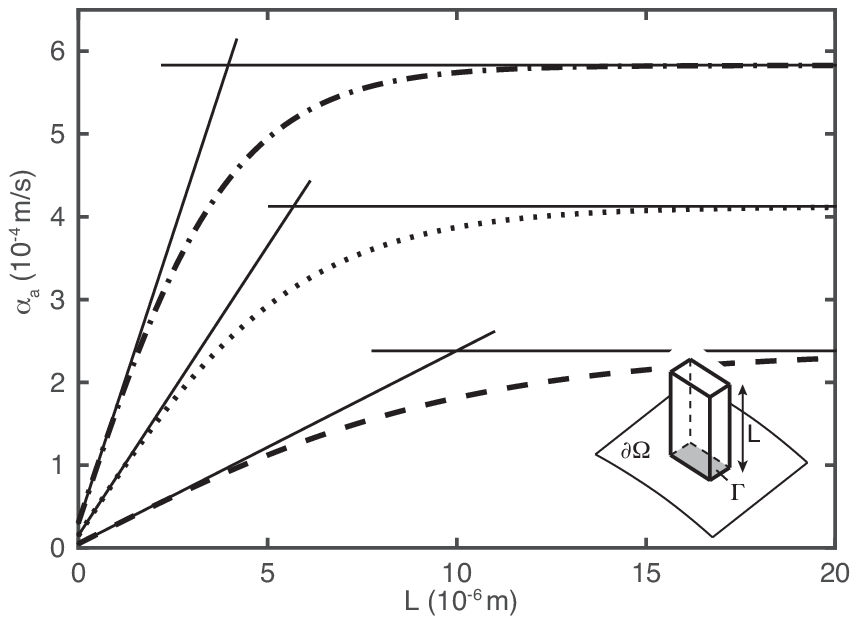}\protect\caption{The average magnetic transfer coefficient $\alpha_{a}$ induced by
a rectangular box with base dimensions $W_{1}=3$ ($10^{-6}$ m), $W_{2}=0.5$
($10^{-6}$ m) and height $L=10$ ($10^{-6}$ m) as a function of
$L$; the base $W_{1}\times W_{2}$ corresponds to the fictitious
boundary ($\Gamma$), the opening to the larger pore. Three different
values of the surface relaxivity were used: $\rho=5$ ($10^{-6}$
m/s) (dashed) $\rho=15$ ($10^{-6}$ m/s) (dotted) and $\rho=30$
($10^{-6}$ m/s) (dashed-dotted). The solid lines depict the initial
slope and the asymptotes.  \label{fig:2}}
\end{figure}
In the case when the domain $\Omega_{1}$ is a separable geometry
and where $\hat{x}$ denotes the normal to $\Gamma$ the Laplace equation
becomes 
\begin{equation}
\begin{cases}
\frac{\partial^{2}H(r)}{\partial x^{2}}=\gamma^{2}H(r) & r\in\Omega_{1}\\
\frac{\partial H(r)}{\partial x}+\rho H(r)=0 & x=L
\end{cases}\label{eq:H_eq}
\end{equation}
 where $\gamma$ denotes the constant of separation. The solution
to equation \ref{eq:H_eq} is then 
\[
H(r)=\sum_{n}^{\infty}B_{n}\left(\frac{(\gamma+\rho)e^{2\gamma L-\gamma x}}{\gamma-\rho}+e^{\gamma x}\right)\phi_{n}
\]
where $\phi_{n}$ denotes the eigenfunctions perpendicular to $x$.
When $\Gamma\rho/D<<1$ (i.e. the spins inside the small domain can
be considered to be in the fast-diffusion regime) the dominant contribution
will be from the lowest relaxation mode $n=0$. Therefore one can
consider the approximate expression 
\[
\alpha_{a}=\frac{1}{\Gamma}\intop_{\Gamma}\frac{\hat{n}\cdot\nabla H(r)}{H(r)}dr\approx\frac{1}{\Gamma}\alpha_{0}\left(\intop_{\Gamma}\phi_{n}d\Gamma\right)^{2}
\]
where the subindex $a$ of $\alpha$ denotes the average magnetic
transfer coefficient and the first coefficient $\alpha_{0}$ is 
\[
\alpha_{0}=\frac{\gamma(\rho\cosh(\gamma L)+\gamma\sinh(\gamma L))}{\rho\sinh(\gamma L)+\gamma\cosh(\gamma L)}.
\]
 A series expansion around $L=0$ gives 
\[
\alpha_{0}=\rho+L\left(\gamma^{2}-\rho^{2}\right)+L^{2}\left(\rho^{3}-\gamma^{2}\rho\right)+O(L^{3})
\]
and for $L>>1$ we reach the asymptotic value 
\[
\alpha_{0}=D\gamma.
\]

We now represent the small geometrical perturbation as a rectangular
box, and show in figure \ref{fig:2} the average magnetic transfer
coefficient $\alpha_{a}$ plotted as a function of the height $L$
of the box where the base area corresponds to the fictitious boundary
$\Gamma$. As the height increases the increased surface area has
less impact on $\alpha_{a}$, which reaches a plateau value. The major
impact comes however from the area of the base, revealed in the separation
constant

\[
\gamma=\sqrt{\rho D\left(\frac{1}{W_{2}}+\frac{1}{W_{1}}\right)}
\]
where $W_{1}$ and $W_{2}$ denote the lengths of the sides of the
base of the rectangular box. While there are sharp corners between
the spherical surface and the rectangular box, we are working with
the Robin kernel, which has a smoothing effect due to the Dirichlet
nature that regularizes this type of sharp features~\cite{Dancer1997}.

As mentioned above, the identified Robin condition in the electrostatic
presentation tends towards a homogeneous boundary condition when $\Omega_{1}\ll\Omega_{0}$.
As a consequence, the original eigenequation describing the spin relaxation
in the total domain $\Omega$ may be well approximated by the altered
(homogeneous) boundary condition of the smooth domain $\Omega_{0}$
in the following way
\[
\begin{cases}
D\Delta u_{n}(r)=\frac{1}{T_{n}}u_{n}(r) & r\in\Omega_{0}\\
(D\hat{n}\cdot\nabla+\rho)u_{n}=0 & r\in\partial\Omega_{0}\\
(D\hat{n}\cdot\nabla+\alpha_{a})u_{n}=0 & r\in\Gamma
\end{cases}.
\]
By adding many small perturbations we reach a rough pore surface with
a significantly increased surface area. This problem may then be simplified
further by defining an effective surface relaxivity $\rho_{e}$ in
the following way
\begin{equation}
\rho_{e}=\frac{\intop_{\partial\Omega}\alpha_{a}dr+\intop_{\partial\Omega}\rho dr}{\intop_{\partial\Omega}dr+\intop_{\partial\Gamma}dr}\label{eq:effective_relaxivity-1}
\end{equation}
and utilizing the well-known analytical solutions for a smooth pore~\cite{brownstein_importance_1979}.
As an example of the usefulness of these results consider a spherical
pore with a radius of $50$ ($10^{-6}$ m) where the sphere surface
is roughened by modulating the surface using small rectangular boxes
(see inset in figure 2). We let the surface relaxivity of the pore
surface (including the surface of the rectangular boxes) be $\rho=5$
($10^{-6}$ m/s) and let the rectangular boxes have the dimensions
$W_{1}=3$ ($10^{-6}$ m), $W_{2}=0.5$ ($10^{-6}$ m) where again
the base area $W_{1}\times W_{2}$ denotes the fictitious boundary
$\Gamma$. Setting the height of the boxes to $L=10$ ($10^{-6}$
m) and using a diffusion constant $D=2$ ($10^{-9}$ $\mbox{m}^{2}$/s)
(corresponding to water at $20$\textdegree{} C) we get the average
magnetic transfer coefficient

\[
\alpha_{a}\approx1.73\mbox{ (}10^{-4}\mbox{ m/s),}
\]
a considerably larger value than the original assigned surface relaxivity
$\rho$. This value describes the average relaxivity over the fictitious
domain $\Gamma$, as induced by the inclusion of the rectangular box
to roughen the surface of the large pore. If we assume that $50$\%
of the spherical surface is covered by such boxes we get the effective
relaxivity

\[
\rho_{e}=1.31\mbox{ }\mbox{ (}10^{-4}\mbox{ m/s}).
\]
The deviance between the original assigned $\rho$ and the effective
$\rho_{e}$ is purely geometrical and due to the roughness of the
pore surface. Given the effective relaxivity it is straightforward
to solve for the lowest relaxation time for this spherical pore~\cite{brownstein_importance_1979}:
\[
T_{0}=0.23\mbox{ s}.
\]
Given the pore size, the original surface relaxivity and the diffusion
coefficient, we would conclude that we would be justified in evaluating
the pore size using the fast-diffusion limit ($R\rho/D\approx0.12$).
This would however give a pore size of $3.45$ ($10^{-6}$ m), a value
which deviates quite substantially from the assigned $R=50$ ($10^{-6}$
m) of the pore we began with. Using instead the effective surface
relaxivity (obtained by calculating the average magnetic transfer
coefficient), we find that the fast-diffusion limit no longer applies
($R\rho_{e}/D\approx3.28$). In order to obtain the correct pore radius
in this regime, one must instead solve for $R$ in the following non-linear
problem (derived by the solution to the Robin problem for a sphere
\cite{brownstein_importance_1979}),
\[
1-\sqrt{\frac{R^{2}}{DT_{0}}}\cot\left(\sqrt{\frac{R^{2}}{DT_{0}}}\right)=\rho_{e}R/D.
\]
This yields $R\approx50.15$ $\mbox{ (}10^{-6}\mbox{ m/s})$, which
is in close agreement with the actual pore radius.

\begin{figure}

\begin{centering}
\includegraphics[width=0.8\columnwidth]{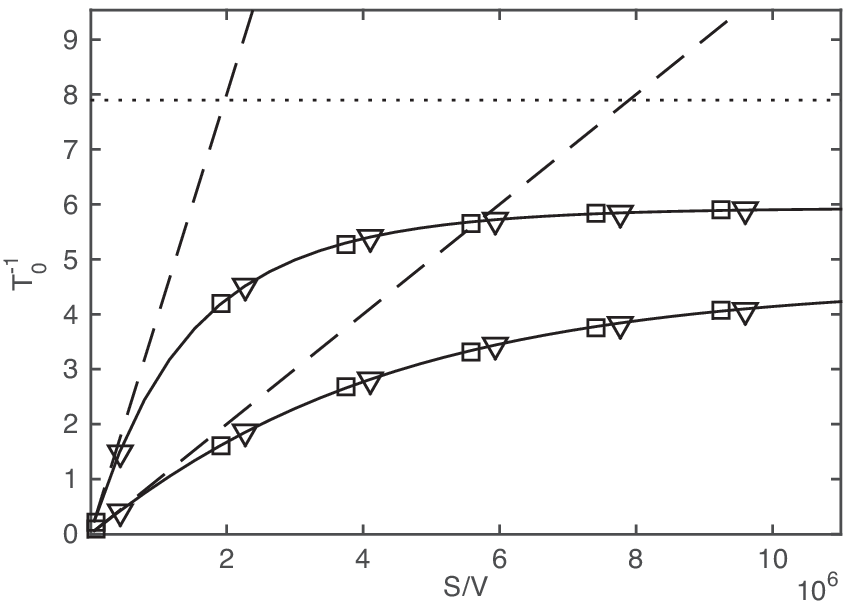}\protect\caption{Inverse relaxation time $T_{0}^{-1}$ as a function of surface-to-volume
($S/V$) for a rough sphere of radius $R=50$ $\mbox{ (}10^{-6}\mbox{ m})$
where the $S/V$ is increased by covering the $50\%$ of the surface
of the sphere by small rectangular boxes of width $W_{1}=W_{2}=0.1$
$\mbox{ (}10^{-6}\mbox{ m})$ and varying the length (squares) and
alternatively covering $50\%$ of the surface by cylinders of radius
$r=0.05$ $\mbox{ (}10^{-6}\mbox{ m})$ and varying the length (triangles).
The dashed line show the fast-diffusion approximation and the dotted
line denote the limit of the slow-diffusion regime. The diffusion
coefficient was set to $D=2$ $\mbox{ (}10^{-9}\mbox{ m}^{2}/s)$
and two values of the surface relaxivity was used, $\rho=1$ $\mbox{ (}10^{-6}\mbox{ m}/s)$
and $\rho=4$ $\mbox{ (}10^{-6}\mbox{ m}/s)$. Both cases are expected
to follow the fast-diffusion limit depicted by the dashed lines and
tend towards the slow-diffusion limit ($\rho R/D>>1$) as $S/V$ increases.
\label{fig:3}}

\par\end{centering}

\end{figure}
The magnetic transfer coefficient is straightforward to obtain analytically
where the surface roughness can be modeled by analytically solvable
geometries. For example a cylinder of radius $R_{C}$ and height $L$
(where the base area $\pi R_{C}^{2}$ corresponds to $\Gamma$) has
the separation constant 
\[
\gamma=\sqrt{\frac{\rho}{D}\frac{1}{r}}
\]
 which coincides with the rectangular box when the $r=W_{1}=W_{2}$,
i.e. the magnetization transfer coefficient becomes equal. The cylinder
gives the initial slope 

\[
\alpha_{a}=\rho+\frac{2\rho}{r}L
\]

and the asymptotic value

\[
\alpha_{a}=\sqrt{2}\sqrt{\frac{\rho D}{r}}
\]
 as $L$ grow large. Figure \ref{fig:3} depicts the inverse of the
relaxation rate of a spherical pore of radius $R=50$ ($10^{-6}$
m) where $50$\% of the spherical surface is covered by such cylinders
as a function of surface-to-volume where the increased surface-to-volume
is modulated by varying the height of the cylinders. Finally, the
developed homogenization procedure has been successfully validated
in numerous numerical examples (not included), with different type
of surface roughness. 

\section{Conclusions}
The presented theory allows us to investigate
the effect of rough surfaces on NMR relaxation experiments by introducing
a magnetic transfer coefficient describing the dissipation of magnetization
over the rough surface. This allows a convenient way of homogenizing
rough pores to smooth equivalents with an effective surface relaxivity.
We find that the effective relaxivity increases dramatically in presence
of rough surfaces, in particular where the surface roughness is narrow
and deep. The explicit expressions for the effective relaxivity demonstrates
the possibility of determining a pore length scale for rough pores when the effective surface relaxivity is known. Our results agrees well with recent experimental findings by Keating~\cite{keating_laboratory_2014}
where surface roughness was induced by etching the surface of glass beads and with numerical simulations by M\"{u}ller-Petke et al.~\cite{muller-petke_nuclear_2015}.

%We would like to thank K. Miller for careful reading of the manuscript
%and acknowledge the financial support by the Knut \& Alice Wallenberg
%foundation.

\end{document}